\documentclass[pra,reprint,twocolumn,showpacs]{revtex4}

\usepackage{amsmath,amsthm,amsfonts,amssymb,bm}
\usepackage{times}
\usepackage{epsfig}
\usepackage[colorlinks={true},dvipdfm]{hyperref}
\hypersetup{citecolor={blue}, filecolor={blue}, linkcolor={blue}, urlcolor={blue}}

\begin{document}
\title{General unifying features of controlled quantum phenomena}
\author{Alexander Pechen}
\author{Constantin Brif}
\altaffiliation{Present address: Sandia National Laboratories,
Livermore, CA 94550, USA.}
\author{Rebing Wu}
\altaffiliation{Present address: Department of Automation and Center
for Quantum Information, Tsinghua University, Beijing 100084,
P.R. China.}
\author{Raj Chakrabarti}
\altaffiliation{Present address: School of Chemical Engineering,
Purdue University, West Lafayette, IN 47907, USA.}
\author{Herschel Rabitz}
\affiliation{Department of Chemistry, Princeton University, Princeton,
New Jersey 08544, USA}

\begin{abstract}
Many proposals have been put forth for controlling quantum phenomena,
including open-loop, adaptive feedback, and real-time feedback
control. Each of these approaches has been viewed as operationally,
and even physically, distinct from the others. This work shows that
all such scenarios inherently share the same fundamental control
features residing in the topology of the landscape relating the target
physical observable to the applied controls. This unified foundation
may provide a basis for development of hybrid control schemes that
would combine the advantages of the existing approaches to achieve the
best overall performance.
\end{abstract}
\maketitle

Steering the dynamics of quantum systems by means of external controls
is a central goal in many areas of modern science, ranging from the
creation of selective molecular transformations to quantum information
processing \cite{Brif2010NJP}. The control may be either coherent
(e.g., a shaped laser pulse~\cite{Ra88-90, OLC, JuRa92, Ra00,
AFC-exp}) or incoherent (e.g., a tailored environment~\cite{ICE} or a
sequence of quantum measurements~\cite{QMC}). Quantum control is
proving to be successful in the laboratory for manipulating a broad
variety of physical, chemical, and biologically relevant
processes~\cite{Brif2010NJP, Ra00, AFC-exp}.

Many seemingly distinct approaches have been developed for controlling
quantum phenomena~\cite{suppl}, including open-loop control (OLC) in
which model-based designs are directly applied in
experiments~\cite{OLC}, adaptive feedback control (AFC) which is a
measurement-guided closed-loop laboratory procedure including
resetting the system to the initial state on each loop
iteration~\cite{JuRa92, Ra00, AFC-exp}, and real-time feedback control
(RTFC) which involves quantum measurement back-action on the system
upon traversing the loop~\cite{Belavkin,Wiseman}. Each of these
control schemes has been argued to be operationally, and even
physically, distinct from the others. This Rapid Communication shows
that all such scenarios are unified on a fundamental level by the
common character of the control landscape~\cite{RHR04-06, Hsieh08}
which relates the physical objective $J[c]$ to the control variables
$c$. This general foundation makes it possible to extend the powerful
landscape-based methods of optimality analysis, originally developed
for OLC and AFC~\cite{RHR04-06}, to RTFC. Furthermore, one can
envision combinations of such schemes \cite{suppl}, or even the
prospect of new unanticipated ones, possibly arising in the future,
all of which will share the same fundamental control landscape
features.

The structure of the control landscape (i.e., its topology
characterized by the nature of the local and global extrema)
determines the efficacy of a search for optimal solutions to the posed
control problem~\cite{Ho06, MHR08}. Searching for the global maximum
of the objective function $J[c]$ in the laboratory would be
significantly hindered by the existence of local maxima that act as
traps during the optimization procedure. Even stochastic optimization
algorithms could be impeded by a high density of traps
\cite{DiMa01}. If the objective function has no traps, then, in
principle, nothing lies in the way of reaching the global maximum
(i.e., the best possible yield), except limited controls. This work
establishes that a common landscape topology is shared by all quantum
control schemes, leading to two general conclusions about controlling
quantum phenomena.  First, there are no local optima to act as traps
on the landscape for a wide class of control problems, thereby
enabling highly efficient searches for globally optimal
controls. Second, for each particular objective $J$, there exists a
single special control $c_*$ which is universally optimal for all
initial states of a system; this result implies that $c_*$ provides
inherent robustness to variations of the initial system's state. For
practical applications, these generic fundamental properties can
facilitate more flexible laboratory implementations of quantum control
as well as guide the design of suitable algorithms for laboratory
optimization~\cite{Roslund09}.

The control landscapes for OLC and AFC were studied in recent works
\cite{RHR04-06, Hsieh08, Ho06, MHR08, OpenSystemLandscape}. For closed
quantum systems, a control objective $J$ can be cast as a function of
unitary evolution operators. Assuming evolution-operator
controllability (i.e., that any unitary evolution can be produced by
the available controls), the landscapes were shown to have no traps
for typical objectives, including the expectation value of an
observable \cite{RHR04-06} and the fidelity of a unitary quantum gate
\cite{Hsieh08}. The absence of landscape traps was also analyzed for
closed systems in the space of actual controls \cite{Ho06}. The
landscapes for OLC and AFC of open quantum systems can be studied by
casting an objective as a function of completely positive,
trace-preserving maps (i.e., Kraus maps \cite{suppl, Kraus83}) which
describe the system evolution. Assuming Kraus-map controllability
(i.e., that any Kraus map can be produced by the available controls),
the trap-free landscape topology was established for open-system
observable control \cite{OpenSystemLandscape}.


The analysis of the control landscapes for RTFC is performed below to
significantly extend the prior analysis~\cite{OpenSystemLandscape} and
provide a single complete framework unifying the landscapes for all
quantum control schemes, including OLC, AFC, and RTFC. Two distinct
approaches to RTFC of quantum systems are considered. The first one is
based on feedback employing measurements of some quantum system output
channel to guide a classical controller \cite{Wiseman, DHJMT00,
RTFC-meas-exp}. The other feedback approach is a quantum analog of
Watt's flyball governor---a self-regulating quantum machine.  In this
scenario (called coherent RTFC), no measurements with a classical
output signal are involved, and instead another quantum subsystem is
employed to facilitate the control action, so that the evolution of
the composite quantum system, consisting of the ``plant'' (the
subsystem of interest) and ``controller'' (the auxiliary subsystem),
is coherent \cite{CohRTFC-theo, CohRTFC-exp}. The performance of
practical RTFC controllers (classical, quantum, or both) can be, in
principle, optimized using AFC \cite{suppl}, possibly guided by an
initial OLC design, further enhancing the significance of a universal
landscape topology common to all quantum control approaches.

In RTFC based on selective measurements of a single quantum system,
the outcome of each measurement is random. After averaging over the
random processes corresponding to all possible measurement outcomes,
feedback control produces Kraus-type evolution of the controlled
system. In RTFC based on measurements of an ensemble of quantum
systems, the ensemble average produces a Kraus map as well. In
coherent RTFC, if the plant and controller are initially prepared in a
product state, the evolution of the plant is once again represented by
a Kraus map. Therefore, all of these incarnations of RTFC generate
Kraus-type evolution. Arbitrary Kraus maps can be engineered, for
example, by using a simple quantum measurement combined with coherent
feedback actions \cite{LlVi01}. Adopting a previous analysis
\cite{OpenSystemLandscape}, we show, under the Kraus-map
controllability assumption, that quantum control landscapes have no
traps for all considered types of RTFC. We will start with basic
definitions, then prove that each considered type of RTFC produces
Kraus-type evolution, and, finally, use these results to determine the
quantum control landscape topology. The conclusion of the Rapid
Communication will then draw the analysis together for a unified
control landscape formulation including OLC, AFC, and RTFC.

In practically important situations the evolution of an open quantum
system can be represented by a Kraus map~\cite{suppl, Kraus83}. For an
$n$-level system, any such map $\Phi$ can be cast as the operator-sum
representation (OSR) \cite{Kraus83}: $\Phi(\rho) = \sum_{\nu=1}^L
K_{\nu} \rho K^{\dagger}_{\nu}$, where $\rho$ is the density matrix (a
positive, unit-trace $n \times n$ matrix) and $\{ K_{\nu}
\}_{\nu=1}^L$ is a set of Kraus operators (complex $n \times n$
matrices satisfying the trace-preserving condition $\sum_{\nu=1}^L
K^{\dagger}_{\nu} K_{\nu} = \mathbb{I}$). The OSR is not unique: Any
map $\Phi$ can be represented using infinitely many sets of Kraus
operators. We denote the set of all Kraus maps by $\mathcal{K}_n$. The
Kraus-map description of open-system dynamics is very general and
includes both Markovian and non-Markovian regimes \cite{LiBiWh01}.

A generalized quantum measurement with $N_0$ possible outcomes
$\{O_{\alpha}\}$ ($\alpha = 1,2,\ldots, N_0$) is characterized by a
family of Kraus operators $\{K_{\alpha,\beta}\}$. We denote the set of
outcomes and corresponding Kraus operators for a given quantum
measurement as $O := \{O_{\alpha} , K_{\alpha,\beta} \}$. In
particular, the projective measurement of a Hermitian observable
$\hat{O} = \sum_{\alpha} O_{\alpha} \Pi_{\alpha}$ with eigenvalues
$O_{\alpha}$ and spectral projectors $\Pi_{\alpha}$ corresponds to the
case $K_{\alpha,\beta} = \Pi_{\alpha} \delta_{\alpha,\beta}$. If the
measurement starts at time $t$ and the system density matrix before
the measurement is $\rho(t)$, then the probability of the measurement
outcome $O_{\alpha}$ will be $p_{\alpha} = \text{Tr} \big[
\sum_{\beta} K_{\alpha,\beta} \rho(t) K^{\dagger}_{\alpha,\beta}
\big]$.

When a selective measurement of duration $\tau_{\text{m}}$ is
performed and the outcome $O_{\alpha}$ is observed, the density matrix
evolves as $\rho(t) \to \rho_{\alpha}(t+\tau_{\text{m}})$, where
\begin{equation}
\label{eq:selective}
\rho_{\alpha}(t+\tau_{\text{m}}) = \frac{1}{p_{\alpha}} \sum_{\beta}
K_{\alpha,\beta} \rho(t) K^{\dagger}_{\alpha,\beta} .
\end{equation} If the measurement is nonselective (i.e., the
measurement outcome is not observed), the corresponding evolution
$\rho(t) \to \rho(t+\tau_{\text{m}})$ will be the average over all
possible measurement outcomes:
\begin{equation}
\label{eq:non-selective}
\rho(t+\tau_{\text{m}}) = \sum_{\alpha} p_{\alpha}
\rho_{\alpha}(t+\tau_{\text{m}}) = \sum_{\alpha,\beta}
K_{\alpha,\beta} \rho(t) K^{\dagger}_{\alpha,\beta} .
\end{equation}
The evolution in Eq.~(\ref{eq:non-selective}) defines a Kraus map
$\Omega_O : \rho(t) \to \Omega_O [\rho(t)]=\rho(t+\tau_{\text{m}})$,
which is completely determined by the measurement $O$.

In RTFC, the measurement (or, in another variation, the interaction
with an auxiliary quantum ``controller'') alters the evolution of the
quantum system at each feedback iteration. Thus, the same quantum
system is followed in real time in the feedback loop. Implementing
RTFC on the atomic or molecular scale appears to be a very challenging
technical problem, but its practical realization promises to
significantly improve the ability to stabilize and control quantum
systems. Consider now RTFC of a single quantum system, where a
discrete series of selective measurements is performed and each
measurement is followed by a feedback action (continuous feedback can
be treated as the limit of the discrete case, resulting in the same
control landscape topology). For the discrete case, the $i$th
iteration of the feedback process consists of a measurement $O^i$ with
possible outcomes $\{O^i_{\alpha}\}$ at time $t_i$, followed by a
feedback action dependent on the measurement outcome (the measured
observables and feedback actions can be distinct at different
iterations). The feedback action may be generally represented by a
Kraus map that depends on the measurement outcome (a special case of
the feedback action is a unitary transformation corresponding to a
coherent control). If at the $i$th iteration the measurement outcome
$O^i_{\alpha}$ is observed, then the feedback action (of duration
$\tau_{\text{f}}$) described by the Kraus map $\Lambda^i_{\alpha}$
will be applied to the system, so that the density matrix will evolve
as $\rho(t_i) \to \rho_{\alpha}(t_i+\tau_{\text{m}}+\tau_{\text{f}})
= \Lambda^i_{\alpha}[\rho_{\alpha}(t_i+\tau_{\text{m}})]$, where
$\rho_{\alpha}(t_i + \tau_{\text{m}})$ is of the
form~(\ref{eq:selective}), with the corresponding probability
$p_{\alpha}^i$. The map $\Lambda^i_{\alpha}$ also includes the free
evolution and influence of the environment. The system evolution after
one feedback iteration, averaged over all possible measurement
outcomes, $\rho(t_i + \tau_{\text{m}} + \tau_{\text{f}}) =
\sum_{\alpha} p_{\alpha}^i
\rho_{\alpha}(t_i+\tau_{\text{m}}+\tau_{\text{f}})$, is therefore
given by the transformation $\rho(t_i) \to \rho(t_i + \tau_{\text{m}}
+ \tau_{\text{f}})  = \Phi^i [\rho(t_i)]$, where
\begin{equation}
\label{eq:one-iteration}
\Phi^i [\rho(t_i)]
= \sum_{\alpha} \Lambda^i_{\alpha} \bigg[ \sum_{\beta}
K^i_{\alpha,\beta} \rho(t_i) (K^i_{\alpha,\beta})^{\dagger} \bigg]
\end{equation}
and $\{ K^i_{\alpha,\beta} \}$ are the Kraus operators characterizing
the measurement $O^i$. Let the OSR of the feedback-action Kraus map
$\Lambda^i_{\alpha}$ be $\Lambda^i_{\alpha}(\rho) = \sum_{\nu}
L^i_{\nu,\alpha} \rho (L^i_{\nu,\alpha})^{\dagger}$, where $\sum_{\nu}
(L^i_{\nu,\alpha})^{\dagger} L^i_{\nu,\alpha} = \mathbb{I}$. Then
Eq.~(\ref{eq:one-iteration}) can be rewritten as
\begin{equation}
\label{eq:Phi-one-iteration}
\Phi^i [\rho(t_i)] = \sum_{\nu,\alpha,\beta} Z^i_{\nu,\alpha,\beta} \rho(t_i)
(Z^i_{\nu,\alpha,\beta})^{\dagger} ,
\end{equation} 
where $Z^i_{\nu,\alpha,\beta} = L^i_{\nu,\alpha}
K^i_{\alpha,\beta}$. Since $\sum_{\nu,\alpha,\beta}
(Z^i_{\nu,\alpha,\beta})^{\dagger} Z^i_{\nu,\alpha,\beta} =
\mathbb{I}$, the transformation $\Phi^i$ of
Eqs.~(\ref{eq:one-iteration}) and (\ref{eq:Phi-one-iteration}) is a
Kraus map. Thus, Eq.~(\ref{eq:Phi-one-iteration}) shows that the
average evolution for one feedback iteration is of the Kraus type for
arbitrary (generalized) quantum measurement and any (coherent or
incoherent) feedback action. For a special case of a pure,
least-disturbing measurement and a coherent feedback action, this
result was obtained in Ref.~\cite{LlVi01}. 

The entire feedback process for controlling the system from the
initial to final time is characterized by a sequence of measurements
and feedback actions: $F=\{O^1,F^1, \ldots, O^N,F^N\}$, where $O^i$
($i = 1,2,\ldots,N$) is the measurement for the $i$th iteration,
$F^i=\{\Lambda^i_{\alpha}\}$ is the set of all feedback actions (Kraus
maps) for the $i$th iteration, and $N$ is the number of
iterations. Different trials of the feedback process will, in general,
produce distinct evolutions, resulting in different system states at
the final time $T$.  
The average output of the feedback process is given by averaging over
all possible evolutions, which produces the transformation $\rho(T)=
\Phi_F [\rho(0)]$, where $\Phi_F = \Phi^N \circ \cdots \circ \Phi^2
\circ \Phi^1$ is a Kraus map given by the composition of one-iteration
Kraus maps of the form (\ref{eq:Phi-one-iteration}).


Consider now RTFC of an ensemble of identical quantum systems, where
measurements record the expectation value $\overline{O} =
\sum_{\alpha} p_{\alpha} O_{\alpha} = \text{Tr} [ \hat{O} \rho(t)]$
of an observable $\hat{O}$. The density matrix representing the state
of the ensemble undergoes a transformation characteristic for a
nonselective measurement (i.e., with averaging over all possible
measurement outcomes): $\rho(t) \to \rho(t+\tau_{\text{m}}) =
\Omega_O [\rho(t)]$, where the Kraus map $\Omega_O$ is defined by
Eq.~(\ref{eq:non-selective}). The feedback action conditioned upon the
measured value $\overline{O}$ is generally represented by a Kraus map
$\Lambda_{\overline{O}}$, so that the ensemble evolution for one
feedback iteration is $\rho(t_i) \to \rho(t_i + \tau_{\text{m}} +
\tau_{\text{f}}) = \Phi^i [\rho(t_i)]$, where the Kraus map $\Phi^i
= \Lambda^i_{\overline{O}} \circ \Omega^i_O$ is the composition of the
Kraus maps representing the ensemble measurement and feedback action
for the $i$th iteration. Similar to the single-system case, the
overall transformation for the entire feedback process is $\rho(T)=
\Phi_F [\rho(0)]$, where the Kraus map $\Phi_F = \Phi^N \circ \cdots
\circ \Phi^2 \circ \Phi^1$ is again the composition of one-iteration
Kraus maps.


Consider now coherent RTFC, where the quantum subsystem of interest
(the plant) interacts with an auxiliary quantum subsystem (the
controller), and the evolution of the composite system is coherent:
$\rho_{\text{tot}}(T) = U(T) \rho_{\text{tot}}(0) U^{\dagger}(T)$.
Here, $\rho_{\text{tot}}$ and $U(T)$ are the density matrix and the
unitary evolution operator, respectively, for the composite system. An
external coherent control field (which is generally time-dependent)
can act on the plant, controller, or both. The state of the plant at
any time $t$ is represented by the reduced density matrix $\rho(t) =
\text{Tr}_{\text{c}}[\rho_{\text{tot}}(t)]$, where
$\text{Tr}_{\text{c}}$ denotes the trace over the controller's degrees
of freedom. If the initial state of the composite system is in tensor
product form, $\rho_{\text{tot}}(0) = \rho(0) \otimes
\rho_{\text{c}}(0)$, then the evolution of the plant is represented by
a Kraus map: $\rho(T) = \Phi_U[\rho(0)] = \sum_{\nu} K^U_{\nu} \rho(0)
(K^U_{\nu})^{\dagger}$,  where the Kraus operators $K^U_{\nu}$ depend
on the evolution operator $U(T)$ of the composite system
\cite{Kraus83}.

With the analysis above, we can now assess the topology of quantum
control landscapes for all considered types of RTFC. To be specific, a
prevalent problem in quantum control is to maximize the expectation
value of some target observable of the controlled system. As we
established above, in measurement-based RTFC for both a single quantum
system and an ensemble, the average density matrix at the final time
$T$ is given by $\rho (T) = \Phi_F [\rho(0)]$, where the Kraus map
$\Phi_F$ is determined by the set $F$ of measurements and feedback
actions at all iterations of the feedback process. The goal is to find
an optimal feedback process $F_{\text{opt}}$ that maximizes the
expectation value of the target observable $\hat{A}$ at the final
time, $\overline{A}(T) = \text{Tr} [\hat{A} \rho(T)]$. The feedback
process $F$ plays the role of a set of controls, and the objective
function has the form: $J[F] = \text{Tr} \big\{\hat{A} \Phi_F[\rho(0)]
\big\}$. An OSR of the Kraus map $\Phi_F$ is given by $\Phi_F
[\rho(0)] = \sum_{\nu} K^F_{\nu} \rho(0) (K_{\nu}^F)^{\dagger}$, where
$\{K^F_{\nu}\}$ is a set of Kraus operators depending on the feedback
process $F$. The objective $J[F]$ can be cast as a function of Kraus
operators:
\begin{equation}
\label{eq:OF-K}
J[\{ K^F_{\nu} \}] = \text{Tr} \left[ \hat{A} 
\textstyle{\sum_{\nu}} K^F_{\nu}
  \rho(0) (K_{\nu}^F)^{\dagger} \right] .
\end{equation} 
We assume that the set $\mathcal{F}$ of all available feedback
processes (i.e., all available measurements and feedback actions) is
rich enough to produce all possible Kraus maps, i.e., that the system
is Kraus-map controllable. For coherent RTFC, the plant also
undergoes a Kraus-type evolution, $\rho(T) = \Phi_U[\rho(0)]$, and the
control objective $\overline{A}(T)$ can also be cast as a function of
the Kraus operators: $J[\{ K^U_{\nu} \}] = \text{Tr} \big[ \hat{A}
\sum_{\nu} K^U_{\nu} \rho(0) (K_{\nu}^U)^{\dagger} \big]$, which has
the same form as Eq.~(\ref{eq:OF-K}). For a sufficiently large
controller 
prepared initially in a pure state, evolution-operator controllability
of the composite system is a sufficient condition for Kraus-map
controllability of the plant \cite{Wu07}.

Building on recent OLC and AFC-motivated analysis of control landscape
topology \cite{OpenSystemLandscape} for a system that is Kraus-map
controllable, the objective function of the form~(\ref{eq:OF-K}) has
no local maxima for any initial state $\rho(0)$ and any Hermitian
operator $\hat{A}$. This result implies that \textit{quantum control
landscapes for OLC, AFC, measurement-based RTFC, coherent RTFC, and
combinations thereof, all have the same trap-free topology}, provided
that the available controls are sufficient to produce any Kraus map.
Furthermore, all local extrema of the objective function of the
form~(\ref{eq:OF-K}) are saddles \cite{OpenSystemLandscape} that can
be easily evaded by a suitable algorithm guiding an ascent over the
landscape. The trap-free control landscape topology is established
above in the space of Kraus maps, $\mathcal{K}_n$. The control
landscape also will be trap-free in the space of actual controls,
$\mathcal{C}$, if the tangent map from $\mathcal{C}$ to
$\mathcal{K}_n$ is surjective everywhere in $\mathcal{C}$. Although,
in general, there exist so-called singular controls~\cite{Bonnard2003}
at which surjectivity does not hold, numerical results show that their
impact on the search for optimal controls should be
negligible~\cite{Wu2009}.

Another important consequence of controllable Kraus-type evolution for
various types of quantum OLC, AFC and RTFC is the existence of a
special control (e.g., for measurement-based RTFC, a special feedback
process $F_{\ast}$) that is optimal for \emph{all} initial states of
the system. Let $\rho(T)$ be the state maximizing the target
expectation value: $\text{Tr}[\hat{A} \rho(T)] = \max_{\rho}
\text{Tr}(\hat{A} \rho)$, and let the spectral decomposition of this
final state be $\rho(T) = \sum_{\alpha=1}^n p_{\alpha}
|u_{\alpha}\rangle \langle u_{\alpha} |$, where $p_{\alpha}$ is the
probability to find the system in the state $|u_{\alpha}\rangle$.  
For an arbitrary orthonormal basis $\{ v_{\beta} \}$ in the system's
Hilbert space, define operators $K_{\alpha,\beta} = p_{\alpha}^{1/2}
|u_{\alpha}\rangle \langle v_{\beta} |$. The Kraus map built from
these operators, $\Phi_{\ast}(\rho) = \sum_{\alpha,\beta=1}^n
K_{\alpha,\beta} \rho K_{\alpha,\beta}^{\dagger}$ generates evolution
$\Phi_{\ast}[\rho(0)] = \rho(T)$ for all initial states $\rho(0)$
\cite{Wu07}. Therefore, the control that produces the Kraus map
$\Phi_{\ast}$ (e.g., for measurement-based RTFC, the feedback process
$F_{\ast} \in \mathcal{F}$ such that $\Phi_{F_{\ast}} = \Phi_{\ast}$)
will be optimal for all initial states and this control will be robust
to variations of the initial system state.

This work shows that the operationally and technologically distinct
quantum control approaches of OLC, AFC, and RTFC share, under the
condition of Kraus-map controllability, a unified control landscape
structure implying two common fundamental properties. First, all such
control schemes are characterized by the absence of landscape traps,
with all local extrema being saddles.  Second, special controls exist
which are universally optimal for all initial states. These findings
establish that (1) there are no inherent landscape features hindering
attainment of the highest possible control yield and (2) suitable
controls can provide broad scale robustness to variations of the
initial conditions. Moreover, these properties are valid for any
quantum control scheme which produces Kraus-type evolution of the
system, including feedback-based and open-loop approaches, as well as
their combinations. The unification of these seemingly different
approaches at a conceptual level (see figures in supplementary
material \cite{suppl}) should, in turn, provide a basis to ultimately
unite the currently distinct laboratory realizations of quantum
feedback control and open-loop control to attain the best performance
under all possible conditions.

This work was supported by NSF and ARO.


\newpage
\onecolumngrid 
\newpage
\setcounter{figure}{0}

\begin{center}
{\large\bf Supplementary material for the manuscript} \\
\vskip 1mm
{\large\bf ``General unifying features of controlled quantum phenomena''}
\vskip 3mm

{\normalsize Alexander Pechen, Constantin Brif, Rebing Wu, Raj
Chakrabarti, and Herschel Rabitz} \\

{\small\it Department of Chemistry, Princeton University, Princeton,
New Jersey 08544, USA}
\end{center}

\setcounter{page}{1}
\thispagestyle{empty}

\section*{Quantum control approaches}

\vspace*{-5mm}
\begin{figure}[ht]
\label{fig1}
\epsfxsize=0.99\textwidth 
\centerline{\epsffile{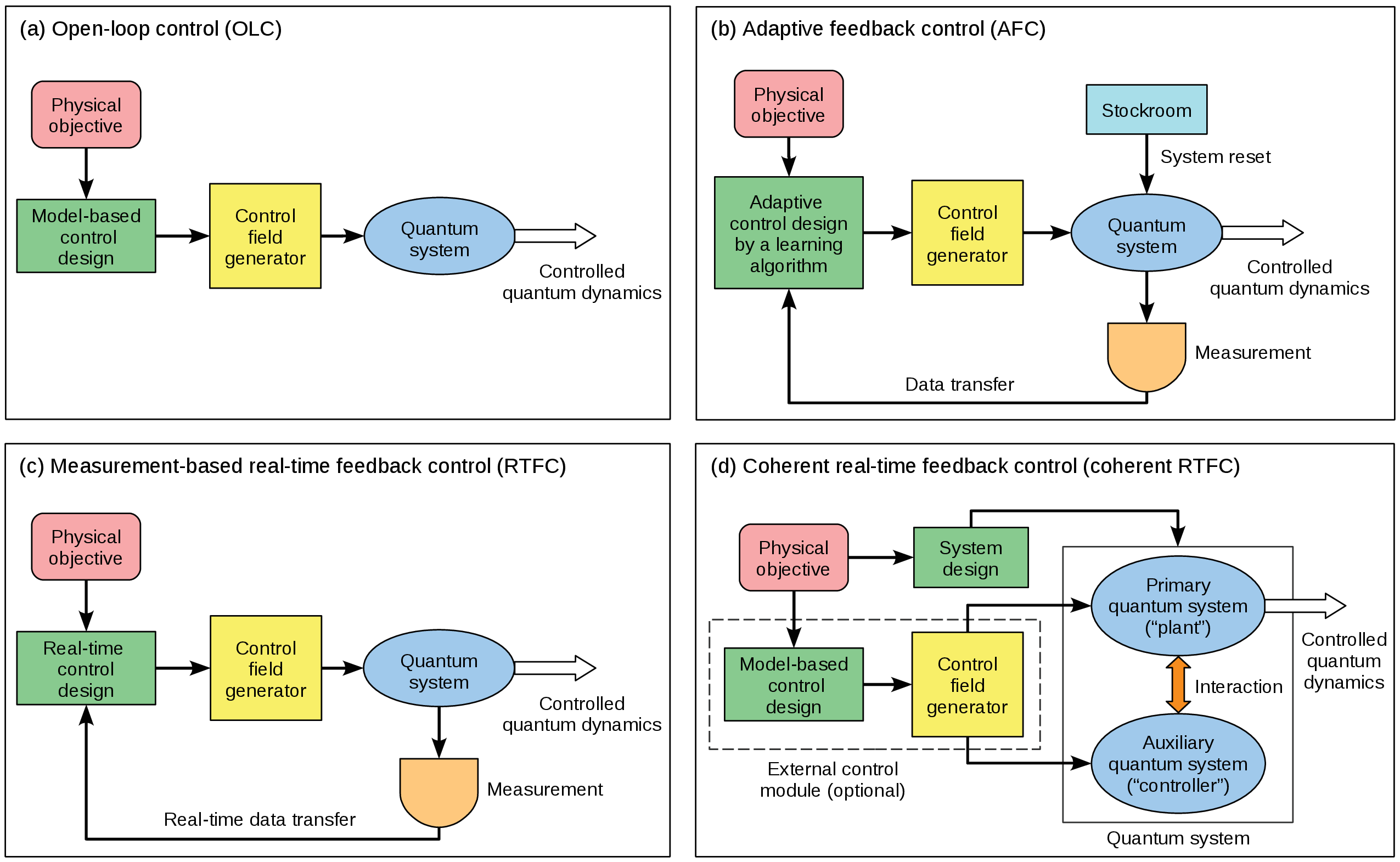}}
\caption{Outline of different quantum control schemes, all of which
aim at achieving a target physical objective by actively manipulating
the dynamics of a quantum system. (a) In open-loop control (OLC), a
theoretical model of the system dynamics is used to design the control
field (e.g., employing optimal control theory). This control design is
then directly applied in the laboratory~\cite{suppl-OLC-exp,
suppl-OLC-rev}. This approach can be satisfactory for some simple
systems; however, in more complex situations, the model-based design
is usually not optimal for the actual
system~\cite{suppl-Brif2010NJP}. (b) In adaptive feedback control
(AFC), the objective is measured at each iteration and the result is
fed back to a learning algorithm that uses previous observations to
propose a new (ideally, better) control~\cite{suppl-JuRa92}.  At the
end of each trial, after the control is applied and the objective is
measured, the system is reset to its initial state for the next
iteration. This resetting makes the back action of the measurement
irrelevant. In recent years, AFC has become a routine procedure that
is successfully used in many laboratories for control of various
quantum phenomena~\cite{suppl-Brif2010NJP, suppl-AFC-exp,
suppl-AFC-rev}. Of particular importance is the ability of AFC to
function in the laboratory in situations when the Hamiltonian of a
complex quantum system and its environment are not well known. (c) In
the measurement-based type of real-time feedback control (RTFC),
measurements are employed to probe the quantum system, and the
obtained information is processed classically in real time to select
the next control action~\cite{suppl-Belavkin, suppl-Wiseman,
suppl-Doherty2000, suppl-WisemanMilburn2010, suppl-MeasRTFC-exp}. In
this scheme, the system evolution during each feedback iteration is
affected not only by the coherent control action (conditioned upon the
measurement outcome), but also by incoherent back-action exerted by
the measurement. Variations of RTFC with both discrete and continuous
measurements were studied \cite{suppl-WisemanMilburn2010}. An
important factor in considering laboratory implementation of RTFC on
the atomic or molecular scale is the issue of loop latency arising due
to the fact that many interesting quantum phenomena occur on a time
scale which is too short to allow for processing of the measured data
in classical controllers based on conventional electronics. (d) In
coherent RTFC, the evolution of the primary quantum system (the
``plant'') is manipulated through its interaction with an auxiliary
quantum system (the ``controller'') \cite{suppl-Lloyd2000,
suppl-James2008, suppl-CohRTFC-exp}. The evolution of the composite
quantum system which consists of the plant and controller is unitary
(assuming that environmental effects are neglected). Optionally, the
plant, controller, or both can be also coherently driven by external
classical control fields (e.g., designed based on a theoretical model
of the composite system dynamics, as in OLC). The optional external
control module is shown within the dashed box. While coherent RTFC can
overcome the latency issue, the quantum controller itself may require
precise engineering to assure quality control performance of the
plant.}
\end{figure}

\section*{A prospective hybrid scheme of quantum control}

\vspace*{-4mm}
\begin{figure}[ht]
\label{fig2}
\epsfxsize=0.83\textwidth 
\centerline{\epsffile{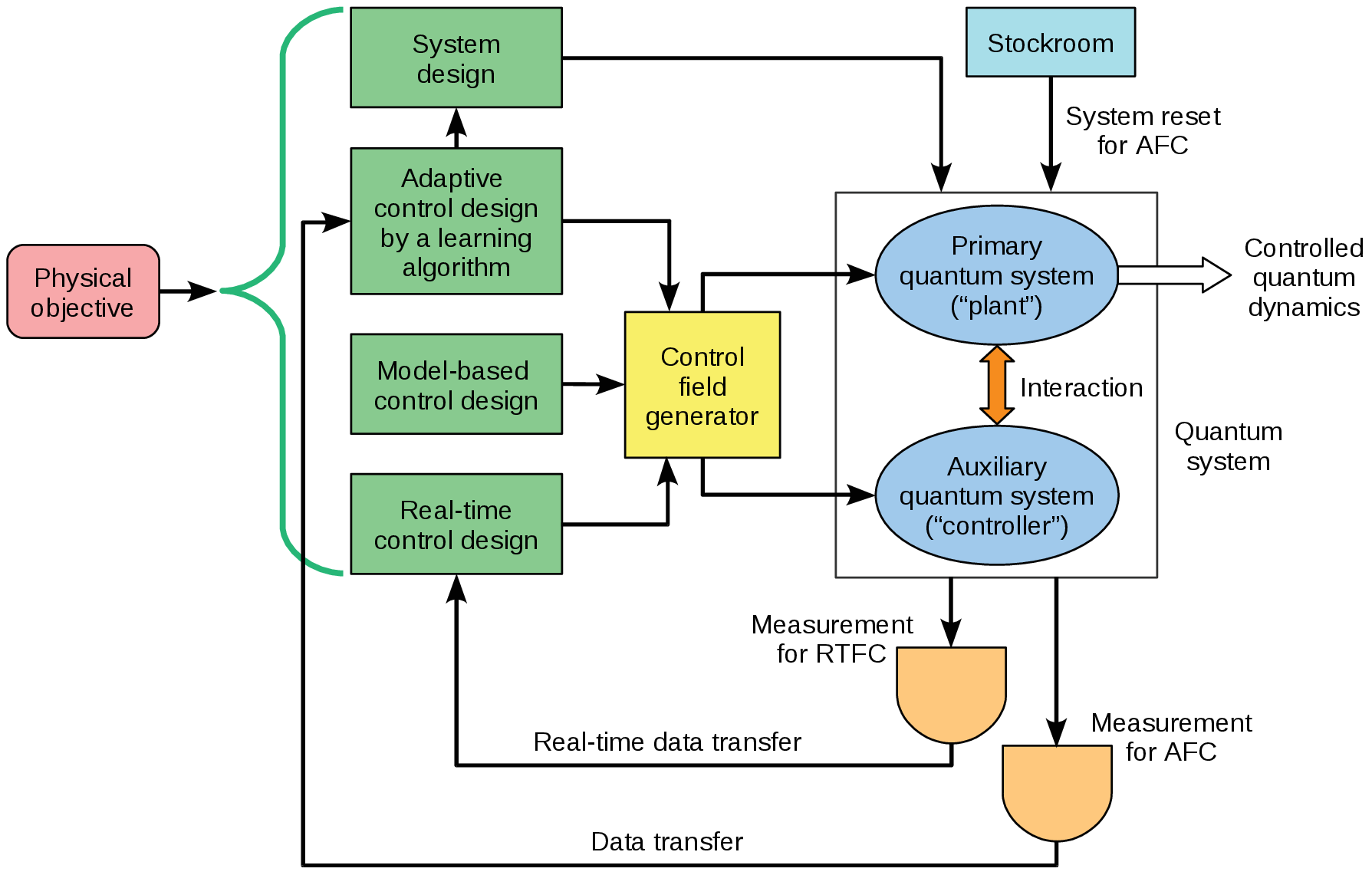}}
\caption{Outline of a prospective hybrid scheme of quantum control
\cite{suppl-Kosut2003}. Since all quantum control approaches,
including OLC, AFC, and different types of RTFC, share the same
trap-free control landscape topology, it would be possible to
consistently maximize the attainment of a target physical objective by
utilizing a hybrid control scheme. Such a hybrid scheme would benefit
from synergetic contributions of all components tuned to achieve the
highest degree of control over the quantum dynamics. In particular, it
can incorporate coherent RTFC which employs the interaction of the
plant (the primary quantum system) with a quantum
controller. Additionally, both the plant and controller can be
manipulated by coherently driving them with external classical control
fields. In order to improve the effectiveness of external controls, a
measurement-based RTFC loop can be implemented, in which the quantum
system is probed at each feedback iteration to assess the best next
control action. Moreover, quantum measurement back-action can be
employed as an additional non-unitary control. The overall performance
of the control system can be optimized in the AFC loop, with the
measurement of the objective followed, at the end of each trial, by
resetting the system to its initial state. AFC can be very useful for
optimizing designs and operational conditions of both quantum and
classical controllers employed in coherent and measurement-based RTFC,
respectively. While model-based theoretical control designs utilized
in OLC will typically not be optimal for actual complex quantum
systems in the laboratory, they can serve as initial trial fields in
the closed-loop optimization as well as play an important role in
exploring the feasibility of various control outcomes. The order, in
which the components of a hybrid control scheme are implemented, can
vary depending on practical circumstances.}
\end{figure}

\section*{Kraus maps}

The density matrix $\rho$ representing the state of an $n$-level
quantum system is a positive, unit-trace $n \times n$ matrix. We
denote by $\mathcal{M}_n =\mathbb{C}^{n\times n}$ the set of all $n
\times n$ complex matrices, and by $\mathcal{D}_n := \{ \rho\in
\mathcal{M}_n \,|\, \rho=\rho^\dagger, \rho \ge 0, \mathrm{Tr}(\rho) =
1\}$ the set of all density matrices. A map $\Phi : \mathcal{M}_n \to
\mathcal{M}_n$ is positive if $\Phi(\rho)\ge 0$ for any $\rho \ge 0$
in $\mathcal{M}_n$. A map $\Phi : \mathcal{M}_n \to \mathcal{M}_n$ is
completely positive (CP) if for any $l \in \mathbb{N}$ the map $\Phi
\otimes \mathbb{I}_l : \mathcal{M}_n \otimes \mathcal{M}_l \to
\mathcal{M}_n \otimes \mathcal{M}_l$ is positive ($\mathbb{I}_l$ is
the identity map in $\mathcal{M}_l$). A CP map is trace-preserving if
$\mathrm{Tr} [\Phi(\rho)] = \mathrm{Tr} (\rho)$ for any $\rho \in
\mathcal{M}_n$. We denote by $\mathcal{K}_n$ the set of all CP,
trace-preserving maps acting in $\mathcal{M}_n$, referred to as
\emph{Kraus maps} or \emph{quantum operations} \cite{suppl-Kraus1983,
suppl-Alicki2007, suppl-Choi1975}.

\end{document}